\documentclass[useAMS,usenatbib,letters]{mn2e}
\usepackage{graphicx}
\usepackage[usenames]{color}
\usepackage{euscript}
\usepackage{journals}

\definecolor{grey}{rgb}{0.5,0.6,0.7}

\title[M59cO -- a missing link between cEs and UCDs]{SDSSJ124155.33+114003.7 -- a Missing Link Between Compact Elliptical and Ultracompact Dwarf Galaxies}
\author[I. V. Chilingarian and
  G. A. Mamon]{Igor V. Chilingarian$^{1,2}$\thanks{E-mail:
    Igor.Chilingarian@obspm.fr} and Gary A. Mamon$^{3}$\\
$^{1}$Observatoire de Paris-Meudon, LERMA, UMR~8112, 61 Av. de
  l'Observatoire, 75014 Paris, France\\
$^{2}$Sternberg Astronomical Institute, Moscow State University, 13 Universitetski prospect, 119992, Moscow, Russia\\
$^{3}$Institut d'Astrophysique de Paris, UMR~7095: CNRS \& 
Universit\'e Pierre et Marie Curie, 98 bis Bd Arago, 75014 Paris, France}
\begin{document}

\date{Accepted 2007 December 17. Received 2007 December 12; in original form 2007 November 18}

\pagerange{\pageref{firstpage}--\pageref{lastpage}} \pubyear{2007}

\maketitle

\label{firstpage}

\begin{abstract}
We report the discovery of a compact object ($R_e = 32$~pc, $M_B =
-12.34$~mag)
 at a projected distance of 9 kpc from 
Messier~59, a giant elliptical in the Virgo
cluster. 
Using HST imaging and
SDSS spectroscopy, both available in the Virtual Observatory, we
find that this object has a blue core containing one-quarter of the light,
and a redder $n=1$ S\'ersic envelope, 
as well as luminosity-weighted age of $9.3\pm1.4$~Gyr, a metallicity
of $-0.03\pm0.04$~dex and a velocity dispersion of $48\pm5$~km$\,$s$^{-1}$.
While ultra-compact dwarfs (UCDs) 
in the face-on view of the Fundamental Plane are found to form a
sequence connecting the highest-luminosity globular clusters with the
lowest-luminosity dwarf ellipticals,  the  compact object near M59 lies in
between this UCD sequence and the positions of compact ellipticals.
Its stellar age, metallicity, and effective surface brightness are similar to 
low-luminosity ellipticals and lenticulars, suggesting that
SDSSJ124155.33+114003.7 is a result of the tidal stripping of such an
object.
\end{abstract}


\section{Introduction}
Compact elliptical (cE) galaxies are high surface brightness, low-luminosity
objects having small effective radii, like M32, the prototypical cE,
are typically an order of magnitude smaller than dwarf elliptical (dE)
galaxies of the same luminosity. But cEs represent a rare class of objects,
including only 6 definite members identified so far. They are found
exclusively in the vicinities of massive galaxies in groups, as M32 and
NGC~5846A, or cluster cD galaxies, as NGC~4486B and
ACO496J043337.35-131520.2 (\citealp{Chilingarian+07}, hereafter C07a), or in
the central regions of massive clusters as two cEs in Abell~1689
\citep{Mieske+05}. The global structural properties of cEs follow those of
giant elliptical galaxies (gEs) and of the bulges of lenticulars and
early-type spirals, placing them on the extension of the Kormendy
(\citeyear{Kormendy77}) relation towards smaller effective radii and higher
surface brightnesses. At the same time, their velocity dispersions are
considerably higher than expected for their luminosities by the
Faber-Jackson (\citeyear{FJ76}) relation.

Galaxies of this class exhibit rather unusual stellar population properties:
their metallicities are much higher than expected for their luminosities;
apart from M32, their stars are usually very old and their $\alpha$/Fe
abundance ratios are significantly super-solar (\citealp{SGCG06};C07a). The
combination of their kinematics and stellar populations supports
the scenario of tidal stripping of more massive galaxies \citep{NP87,CGJ02}. 

Ultra-compact dwarf galaxies (UCDs, \citealp{PDGJ01}) initially discovered
as extragalactic sources unresolved from the
ground-based observations, represent another class of compact stellar
systems.
At least an order of magnitude
smaller than cEs ($R_{e}\sim$10~pc), but still significantly larger than
globular clusters (GCs, \citealp{Drinkwater+03,Jordan+05}), UCDs are well
studied only in the two nearest clusters of galaxies: Fornax
\citep{Drinkwater+00} and Virgo \citep{Hasegan+05,Jones+06}. Some candidate
UCDs have been also found in nearby galaxy groups
\citep{Evstigneeva+07}.

UCDs, initially defined on a morphological basis, probably represent a
heterogeneous class of objects \citep{MHIJ06} of different origins.
The concepts of UCD formation include: (1) 
very massive globular clusters having the same origin as ``normal'' ones
\citep{MHI02}; (2) stellar superclusters formed in
gas-rich mergers of galaxies \citep{FK02,FK05}; (3) 
end-products of small-scale primordial density fluctuations in
dense environments \citep{PDGJ01}; (4)
tidally stripped nucleated dEs (dEN's, \citealp{BCD01,BCDS03})
or simply dEN's with very low surface brightness outer components 
(cf. \citealp{Drinkwater+03}).

UCDs occupy a region of the fundamental plane (FP, \citealp{DD87}) between
the sequences of globular clusters and dEs \citep{Drinkwater+03}. Up-to now,
the structural, dynamical and stellar population properties of the brightest
and most massive UCDs are still quite far from those of cEs.

In this \textit{Letter} we report the discovery of a compact galaxy in the
Virgo cluster with the properties placing it between the brightest UCDs and
the cEs. Our results are based on the analysis of archival spectral and
imaging data, available in the International Virtual Observatory.

\section{Discovery of SDSSJ124155.33+114003.7: Data and Techniques Used}
SDSSJ124155.33+114003.7 (hereafter, M59cO for compact object) has been
serendipitously
discovered in SDSS-DR6  \citep{SDSS_DR6} in
the course of a study of the luminosity function of the Virgo cluster (Mamon
et al. in prep.): M59cO is the only object within 
62~deg$^{2}$ ($1 R_{\rm vir}$) around M87, classified as a galaxy by the SDSS 
photometric and spectroscopic pipelines but as
a point source visually. It lies just 2.1~arcmin to the Northwest of
the 
giant
elliptical galaxy M59, which itself sits in the nearest known compact group
of galaxies \citep{Mamon89}, which contains the even brighter galaxy M60. We
adopt the M59 distance modulus of 30.87~mag \citep{Mei+07}, yielding
a spatial scale of 72~pc~arcsec$^{-1}$.

We have searched for the complimentary datasets in the Virtual Observatory
using CDS Aladin \citep{Bonnarel+00}, and found: (1) a spectrum ($R$=1800)
of this galaxy in a 3~arcsec-wide aperture obtained by the SDSS; (2) fully
calibrated HST ACS (WFC) images in the two photometric bands, F475W ($g'$)
and F850LP ($i'$) from the Virgo Cluster ACS Survey \citep{Cote+04} provided
by the ACS Associations service of the Canadian Astronomical Data Center.

The high resolution evolutionary synthesis simple stellar populations,
computed with {\sc PEGASE-HR} \citep{LeBorgne+04}, have been fit against the
SDSS-DR6 spectrum of M59cO in order to obtain parameters of the stellar
population and internal kinematics, using the ``NBursts'' spectral fitting
technique \citep{CPSA07} for a single Simple Stellar Population (SSP).  The
variations of the spectral resolution as a function of the wavelength
provided by the SDSS were used to broaden the SSP models. We have used the
wavelength range between 4050 and 5700\AA. The blue limit is due to the
wavelength range of the {\sc PEGASE-HR} models (4000\AA), while the red
limit has been set to work only in the blue arm of SDSS and avoid the
``jump'' of the line-spread-function width in the combined SDSS spectrum.

\section{Properties of SDSSJ124155.33+114003.7}

\subsection{Photometric properties}

\begin{figure}
\includegraphics[width=0.49\hsize]{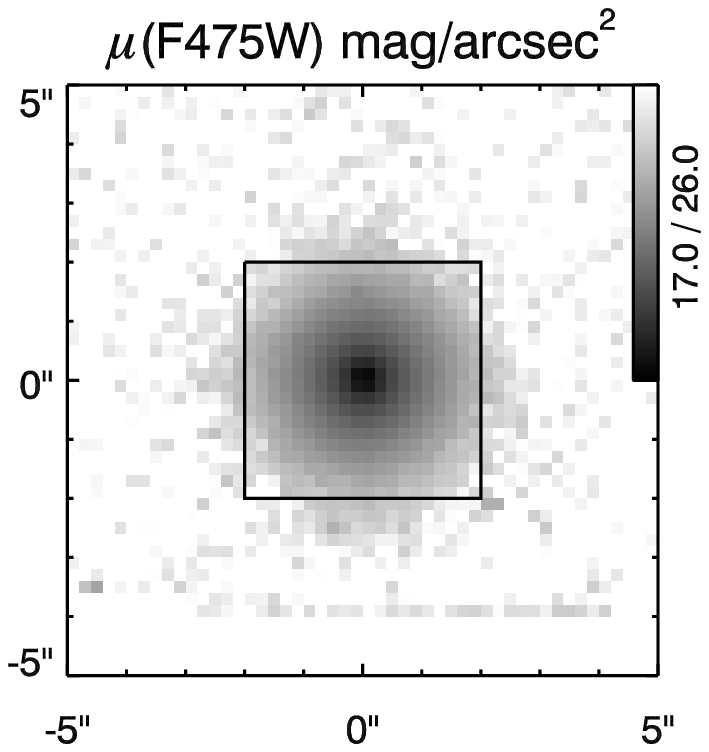} 
\includegraphics[width=0.49\hsize]{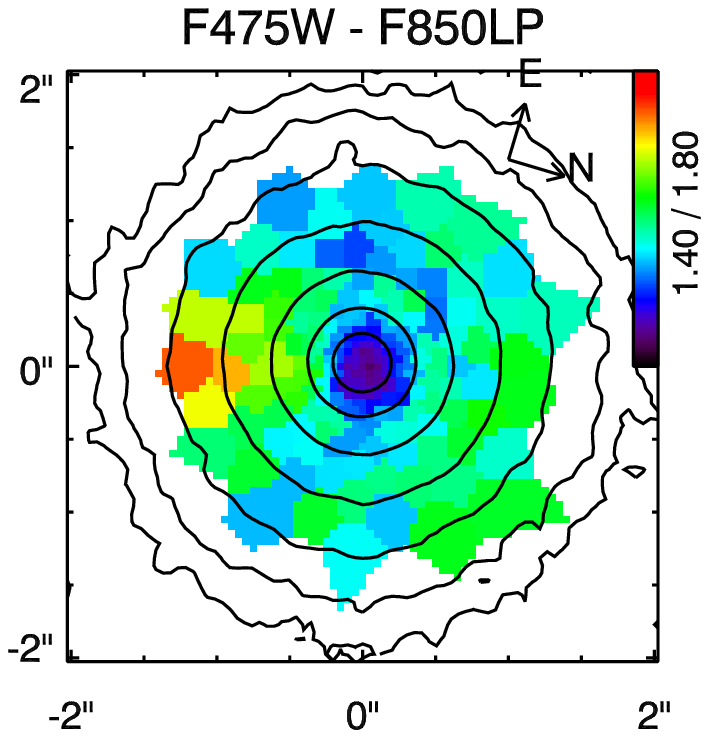} 
\includegraphics[width=8.0cm]{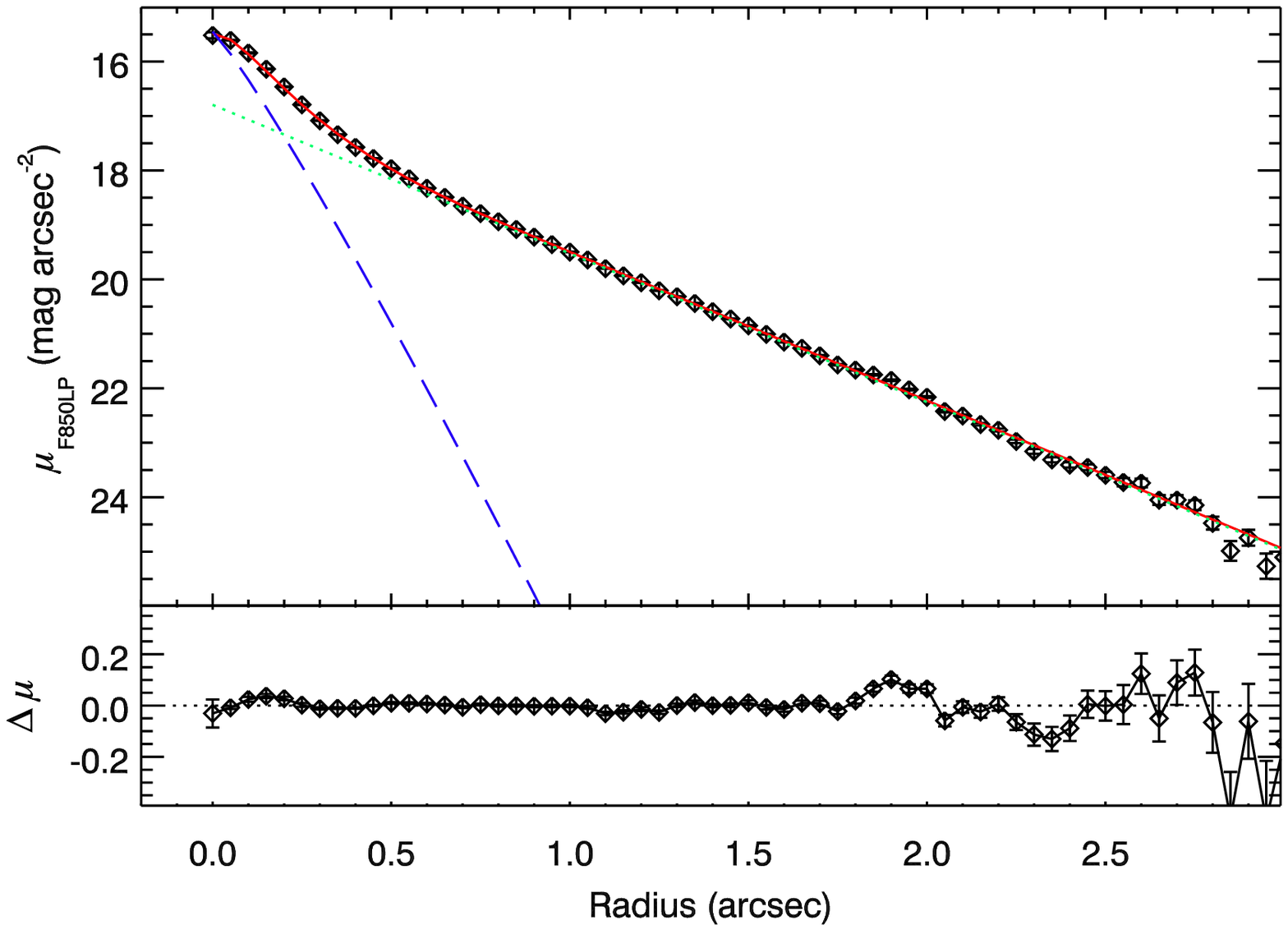} \\
\caption{\emph{Top left}: HST ACS image of M59cO in the $g'$ band;
\emph{Top right}: $g'$--$i'$ colour map, derived from the HST ACS; 
\emph{Bottom}: light profile of
M59cO in the F850LP photometric band with its best-fitting
model, containing an inner S\'ersic core with $n$ close to 1 (\emph{blue
dashed line}), and an exponential outer profile (\emph{green dotted line}) and
their superposition \emph{red solid curve}.
\label{figprof}}
\end{figure}

The total (corrected isophotal) F475W AB magnitude ($m_{F475}$=$18.12$$\pm$$
0.03$), measured using {\sc SExtra\-ctor} \citep{BA96}, is in good agreement
with the SDSS values ($m_{g'}$=$18.08$$\pm$$0.01$, Petrosian magnitude). 
Converting this magnitude into the $B$ band according to \cite{FSI95},
assuming an elliptical galaxy SED ($B$-$g'$ = 0.55~mag)\footnote{The stellar
population parameters and broad-band colours of M59cO are similar to those
of intermediate-luminosity ellipticals.} and correcting it for the Galactic
extinction \citep{SFD98}, we end up with a total absolute magnitude
$M_B=-12.34$. With {\sc SExtractor}, we find an effective radius of
$R_{e}$=0.43~arcsec$\equiv$32~pc, yielding a mean surface magnitude within
$R_e$ of $\langle\mu_B\rangle_e$=18.69$\,$mag$\,$arcsec$^{-2}$, as bright as
the brightest known UCDs.\footnote{Two objects from \cite{EGDH07}, VUCD~7
and UCD~3 are slightly more luminous. However, both of them have extended
``halos'', containing a significant fraction of the stellar light. So the
recovered luminosities of the compact components, derived from their
Table~10, are considerably lower than that of M59cO. On the other hand, the
faintest known cE, M32, is about 15 times more luminous ($M_B=-15.34$,
\citealp{Graham02}, converted into the $B$ band).}

The unsharp masking technique, applied to the images, has not revealed any
structures embedded in the object. We have fit the HST/ACS surface
photometry of M59cO using both the IRAF {\sf stsdas.isophote.ellipse} task
and the {\sc GalFit} software package \citep{PHIR02}, see Fig.~\ref{figprof}
(bottom). Both techniques give highly consistent results. The model
profiles have been convolved with the ACS PSF constructed for both filters at
the position of M59cE on the ACS frame using the {\sc Tiny Tim}
software\footnote{http://www.stsci.edu/software/tinytim/} with the K4V star
spectrum as a template. None of the S\'ersic, King, nor `Nuker'
models was able to represent well the light distribution of the galaxy. The
best 2-component fitting is obtained with either two S\'ersic profiles with
$n \simeq 1$ (i.e. exponential profiles) or else an exponential profile and
a compact King core. The galaxy is almost perfectly round --- the
ellipticity of the isophotes remains under 0.03 at all radii. The $B$-band
extinction-corrected best fitting values for the inner and outer components
are as follows: $R_{e,\mbox{in}} = 13$$\pm$1~pc,
$\langle\mu\rangle_{e,\mbox{in}} = 18.13$$\pm$$0.09$~mag~arcsec$^{-2}$,
$m_{\mbox{tot},\mbox{in}} = 19.78$~mag, $R_{e,\mbox{out}} = 50 \pm 2$~pc,
$\langle\mu\rangle_{e,\mbox{out}} = 20.01$$\pm$$0.04$~mag~arcsec$^{-2}$,
$m_{\mbox{tot},\mbox{out}} = 18.82$~mag.

We obtain a $g'$--$i'$ colour map by  degrading
the F475W image, which has slightly better resolution (FWHM=0.090~arcsec), 
to the resolution of the F850LP image (0.092~arcsec) 
using the simple transformation: $I_{\mbox{475}}^{\rm
  degraded} = 
  F^{-1}[F(I_{\mbox{475}})\cdot F({\rm PSF}_{\mbox{850}}) / F({\rm
  PSF}_{\mbox{475}})]$, 
  where $F$ and $F^{-1}$ stand for direct and inverse Fourier transforms,
  respectively.
For
colour information at the periphery of the object, we have applied Voronoi
adaptive binning \citep{CC03} to the F850LP image with the target
signal-to-noise ratio of 100.

The $g'$--$i'$ colour map obtained in this fashion is presented in
Fig.~\ref{figprof} (upper right).  Tessellae containing more than 120 pixels
(low signal regions) are masked.  The core is $\sim$0.15 magnitude bluer
than the surrounding envelope. The size of the blue core matches well the
parameters of the inner component (exponential or King core) obtained by
{\sc GalFit} and 1D light profile analyses. It may be the signature of a
younger stellar population in the nucleus of M59cO, as found in three bright
dEs in Virgo and a low-luminosity lenticular NGC~130 \citep{CSAP07,CSAP07b}.

\subsection{Kinematics and Stellar Population}

\begin{figure}
\includegraphics[width=\hsize]{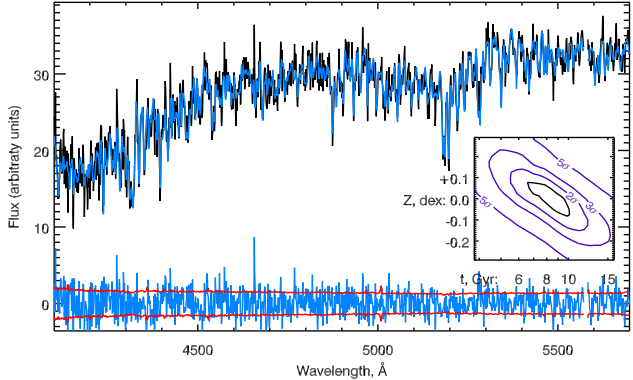}
\caption{SDSS spectrum of M59cO, its best-fitting
template and the residuals of the fitting; 
\emph{Inner panel}: 1 (black), 2, 3, and $5\,\sigma$ confidence levels
in age-metallicity space.
\label{figspec}}
\end{figure}

Fitting the SDSS spectrum of M59cO,
we have derived the following values of kinematical and stellar
population parameters: $v = 708$$\pm$3~km$\,$s$^{-1}$, 
$\sigma_v = 48$$\pm$5~km$\,$s$^{-1}$, $t = 9.3$$\pm$1.4~Gyr, 
$Z = -0.03$$\pm$0.04~dex. The spectrum and its best-fitting SSP model are
shown in Fig.~\ref{figspec}. The radial velocity of M59cO
differs by $\sim$270~km$\,$s$^{-1}$ from M59 ($v_{r} = 440$~km$\,$s$^{-1}$). This
can be considered as evidence of their gravitational interaction.
We scanned the $\chi^2$ parameter space on a pre-defined age-metallicity
grid as described in Appendix~A of \cite{CPSA07}, to provide
realistic uncertainties of the stellar population parameters and locate
possible secondary minima. The confidence contours in the age-metallicity
space are shown in Fig.~\ref{figspec} (inset).

We have calculated the Lick indices \citep{WFGB94} and derived
the [Mg/Fe] abundance ratio using models of \cite{TMB03} as
$+0.21\pm0.10$~dex. However, the fitting residuals do not show the
characteristic ``step'' at 5185\AA\ usually seen for super-solar
[Mg/Fe] abundance ratios due to template mismatch, thus we suspect some
contamination of the Fe$_{5270}$ feature, perhaps by a faint cosmic ray hit,
biasing the measurements of the corresponding Lick index. Thus, we consider
the value of $+0.2$ as an upper limit to the [$\alpha$/Fe] abundance
ratio.

\section{Discussion and conclusions}

\subsection{Comparison with known UCDs and cEs}

\begin{figure}
\includegraphics[width=8cm]{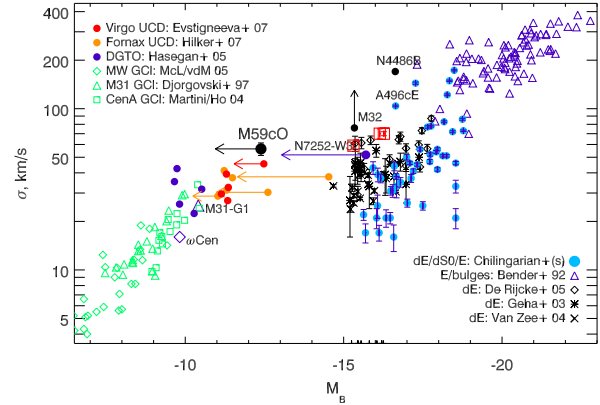}
\caption{Faber-Jackson relation for giant and dwarf early-type galaxies, 
bulges of spirals, UCDs, globular clusters and transitional objects. The 
sources of the data are described in the text. Red squares mark 
NGC~4467, IC~3653, and VCC~1627 (see Section 4.2 for details).
\label{figfjr}}
\end{figure}

\nocite{Chil07A496} 
\nocite{EGDH07}
\nocite{Hilker+07}
\nocite{Hasegan+05}
\nocite{MvdM05}
\nocite{Djorgovski+97}
\nocite{MH04}
\nocite{Maraston+04}
\nocite{vZSH04}
\nocite{deRijcke+05}
\nocite{GGvdM03}

Fig.~\ref{figfjr} presents the Faber-Jackson (\citeyear{FJ76}) relation for
dynamically hot stellar systems. It is an extended version of Fig.~3 from
C07a, and includes a full sample of Abell~496
galaxies, 
UCDs in Virgo
 and Fornax; dwarf-globular
transition objects (DGTOs); GCs in
the Milky Way, M31, and
NGC~5128; and W3-NGC~7252, the ``heavy-weight GC''. 
The velocity dispersions of M59cO and W3 have
been corrected by a factor of 1.14 
to correct from global to central values
\citep{Djorgovski+97}.
Other sources of data remain the same as in
Fig.~3 in C07a.
The three UCDs in Fornax and one in Virgo, having extended halos, are shown as
arrows with ends (as is M59cO), 
corresponding to the parameters of the compact central
structures. For W3, the left end of the arrow,
corresponds to the predicted luminosity of this
object at the age of 10~Gyr (data from \citealp{Maraston+04}).

On the Faber-Jackson diagram, the newly discovered object lies on the bright
end of the sequence formed by globular clusters, DGTOs and UCDs, being
almost two magnitudes brighter. A few objects, including all known cEs,
reside on the continuation of this sequence at higher
luminosities, exhibiting velocity dispersions about three times higher than
dE/dS0 galaxies of the same brightness.

As already mentioned, the three UCD galaxies with extended stellar halos (2
in Fornax and 1 in Virgo) brighter than M59cO, move left in Fig.~\ref{figfjr} to
the ``usual'' locus of the UCDs, if one only considers their compact
components (ends of the arrows in the figure). 
At the same time, the predicted position of the young heavy-mass cluster
NGC7252-W3, after 10~Gyr of passive evolution, is very close to
M59cO.

\begin{figure}
\includegraphics[width=7.5cm]{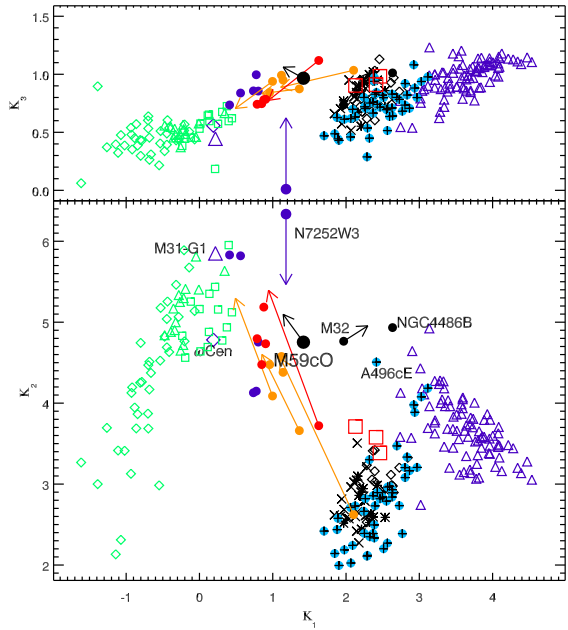}
\caption{Fundamental plane (in $\kappa$-space) for different classes
  of round stellar systems.
Same 
symbols and colours as in Fig.~\ref{figfjr}. \label{figfpk}}
\end{figure}

A more detailed comparison of the physical properties of M59cO with other
types of galaxies and GCs is given by the Fundamental Plane (FP).  We use
the redefinition of the FP in the $\kappa$-space proposed by \cite{BBF92},
where $\kappa_1$ is related to the logarithm of the total mass,
$\kappa_2$ proportional to the $(M/L)I_e^3$ is a measure of ``compactness'',
and $\kappa_3$ is connected to the logarithm of the mass-to-light ratio.

In Fig.~\ref{figfpk} we show the two views of the fundamental plane:
$\kappa_3$~vs~$\kappa_1$ (``edge-on'') and $\kappa_2$~vs~$\kappa_1$ 
(``face-on''). The regions, occupied by the different classes of objects,
mentioned above, are distinctly visible on the ``face-on'' view, while the
``edge-on'' one demonstrates only two sequences: (1) dEs, cEs, gEs and the
bulges of spirals on the right and (2) GCs, DGTOs and UCDs on the left. The
``face-on'' (compactness vs mass) view of the FP shows that \emph{UCDs form
a sequence connecting the high-luminosity GCs with the low-luminosity dEs}.
Despite its two-component structure, the outer component of M59cO is still
more compact than the extended halos of UCD~3, UCD~5, FCC~303, and VUCD~7
from \cite{EGDH07}, whose $R_{e}$ range from 107~pc (UCD~3) to 696~pc
(FCC~303) and whose quite low surface brightnesses
($\langle\mu_B\rangle_{e}$ from 21.1 to 23.2~mag~arcsec$^{-2}$), make them
rather regular structures, reminiscent of normal dEs.

On the FP, M59cO lies between the loci of UCDs and cEs-gEs, still being
closer to UCDs. However, since its stellar population is similar to that of
M32 \citep{Rose+05}, we conclude that \emph{SDSSJ124155.33+114003.7 is a
transitional object between UCDs and compact elliptical galaxies}.

\subsection{Origin of SDSSJ124155.33+114003.7}
The dynamical mass-to-light ratios of UCDs vary quite significantly 
\citep{Drinkwater+03,Hasegan+05,Hilker+07}, suggesting the presence of
dark matter (DM) in some of them. \cite{Hasegan+05} propose to use $M/L$
ratio as a criterion to distinguish between UCDs and massive
GCs. At the same time, \cite{EGDH07} find no evidence for dark
matter in UCDs.

The virial theorem mass estimate is $M_{\rm vir}
\sim 10.0 R_{e} \sigma_v^2 / G$, where 
$\sigma_v$ is the global velocity dispersion (\citealp{Spitzer69}, with the
correction from 3D to 2D half-light radius as in the \citealp{Hernquist90}
model). The total mass of M59cO is then $(1.5 \pm 0.3) \times 10^8
M_{\odot}$, in between the most massive known UCD and the least massive
known cE.  The $B$-band mass-to-light ratio of the stellar population,
estimated from the {\sc PEGASE-HR} evolutionary synthesis models for Salpeter
IMF ($5.9 \pm 0.7$) translates into the stellar mass $M_{*} = (9.1 \pm 1.7)
\times 10^7 M_{\odot}$, suggesting that M59cO contains about 40 per cent of
DM, making it similar to much more massive dE satellites of M31 
\citep{DRPSD06}. Had we adopted a \cite{Scalo86} or \cite{Kroupa01} IMF, 
we would have found less stellar mass, hence a higher fraction of DM.

Given the high metallicity of its stellar population, its age of only
$\sim$9~Gyr suggests that M59cO cannot be a passively evolved primordial
object.  The presence of DM allows to conclude that it is neither
an enormous globular cluster, nor could it have formed as a tidal object in
the process of galaxy merger, as the NGC~7252-W3 supercluster was probably
created. So the most tempting possibility is that \emph{M59cO is the tight
core of a larger galaxy, severely stripped by the tidal field of M59}.

Comparing the age and metallicity of M59cO to those of dEs and
intermediate-luminosity Es \citep{SGCG06,Chil07A496}, we estimate the
luminosity of its progenitor to be in the range $-16.0 < M_B < -18.0$. 
Another argument for this scenario is the relatively high surface brightness
of the outer component of M59cO: its progenitor could not be a faint dE,
otherwise one would expect a much fainter ``halo'' similar to those of the
four UCDs in \cite{EGDH07}.

Six early-type galaxies from the Virgo ACS Survey host stellar nuclei
comparable in luminosity and half-light radii to the inner component of
M59cO. We have estimated kinematical and stellar population parameters for
four of them, NGC~4387, IC~3653, IC~3328, VCC~1627, having SDSS-DR6 spectra,
while for the remaining two, NGC~4379 and NGC~4467, the literature data have
been used \citep{Sil06,Trager+98}. All these galaxies (except IC~3328, which
is younger and more metal-poor) exhibit metallicities between $-0.1$ and
+0.15~dex and ages between 5 (IC~3653) and 12 (NGC~4379) Gyr, i.e. similar
to M59cO. However, the [Mg/Fe] abundance ratio is strongly super-solar for
NGC~4379, and velocity dispersions of NGC~4379 and NGC~4387 are above
100~km$\,$s$^{-1}$. The remaining three galaxies with $-17.0 < M_B < -16.0$
are rather bright and compact and can be
considered as ``transitional'' objects between dE/dS0 and E/S0. 
The surface brightness of the inner regions is
similar to that of the outer component of M59cO.

Therefore, our best explanation is that \emph{SDSSJ124155.33+114003.7 is a
result of the tidal stripping of a transitional object between dwarf and
normal early-type galaxies}, known to have higher surface brightness than
fainter dE/dS0s or brighter gEs. In this case, the outer component of M59cO can be
considered as a dynamically heated remnant of the progenitor's thick disc.

The properties of the newly discovered object do not allow to classify it
uniquely. It can be considered either as the faintest cE, or as the
brightest UCD. However, since it is a product of a violent tidal stripping,
as cEs and at least some UCDs are, the exact classification is not that
important. The discovery of M59cO contributes a useful piece to the 
puzzle of galaxy evolution in dense environments.

\section*{Acknowledgments}
We thank E. Evstigneeva for providing a compilation of data for 
globular clusters and UCDs in computer-readable form;
V\'eronique Cayatte, Sven De Rijcke and Simona Mei for fruitful discussions;
and our referee, Michael Drinkwater, for his concise 
and clear report and useful comments; VO Paris 
Data Centre for computational resources.

\bibliographystyle{mn2e}
\bibliography{cOM59}

\label{lastpage}

\end{document}